\documentclass{IEEE_lsens}
\usepackage{textcomp}

\usepackage[noadjust]{cite}
\usepackage[T1]{fontenc} 
\usepackage{amsmath}

\usepackage[]{graphicx}
\usepackage[varg]{txfonts}
\usepackage[T1]{fontenc} 
\usepackage{subfigure}
\usepackage{subfigmat}
\usepackage{algorithm}
\usepackage{nomencl}
\usepackage{color}
\usepackage{soul}
\usepackage{algpseudocode}
\usepackage{epstopdf}
\newcommand{\argmax}{\mathop{\rm arg~max}\limits}

\usepackage[cmintegrals]{newtxmath}
\usepackage{bm}
\usepackage{array}
\usepackage{url}

\providecommand{\hypersetup}[1]{\relax}

\begin{document}
\markboth{Vol., No., 2020}{0000000}
\IEEELSENSarticlesubject{Sensor signal processing}

\title{Data-driven Vector-measurement-sensor Selection based on Greedy Algorithm}

\author{\IEEEauthorblockN{Yuji~Saito\IEEEauthorrefmark{1}\IEEEauthorieeemembermark{1}, Taku~Nonomura\IEEEauthorrefmark{1}\IEEEauthorieeemembermark{1}, Koki~Nankai\IEEEauthorrefmark{1}\IEEEauthorieeemembermark{1},~Keigo~Yamada\IEEEauthorrefmark{1}\IEEEauthorieeemembermark{1}~and~Keisuke~Asai\IEEEauthorrefmark{1}\IEEEauthorieeemembermark{1}}
\IEEEauthorblockA{\IEEEauthorrefmark{1}Department of Aerospace and Engineering, Tohoku University, Miyagi 980-8579, Japan}
\IEEEauthorblockN{Yasuo~Sasaki\IEEEauthorrefmark{2}\IEEEauthorieeemembermark{1},~and~Daisuke~Tsubakino\IEEEauthorrefmark{2}\IEEEauthorieeemembermark{2}}
\IEEEauthorblockA{\IEEEauthorrefmark{2}Department of Aerospace and Engineering, Nagoya University, Nagoya, Aichi 464-8603, Japan\\
\IEEEauthorieeemembermark{1}Non-Member, IEEE\\
\IEEEauthorieeemembermark{2}Member, IEEE}%

\thanks{Corresponding author: Yuji Saito (e-mail: saito@aero.mech.tohoku.ac.jp).\protect\\
}
\thanks{Associate Editor: }%
\thanks{Digital Object Identifier }}

\IEEELSENSmanuscriptreceived{Manuscript received , 2020;
revised , 2020; accepted , 2020.
Date of publication , 2020; date of current version , 2020.}

\IEEEtitleabstractindextext{%
\begin{abstract}
A vector-measurement-sensor problem for the least squares estimation is considered, by extending a previous novel approach in this paper. An extension of the vector-measurement-sensor selection of the greedy algorithm is proposed and is applied to particle-image-velocimetry data to reconstruct the full state based on the information given by sparse vector-measurement sensors.
\end{abstract}

}

\maketitle
\section{Introduction}
\label{sec:intro}
Reduced-order modeling for fluid analysis and flow control gathers a lot of attention because of a short time processing for a large amount of analysis data and the stability of real-time feedback control. With regard to reduced-order modeling, proper orthogonal decomposition (POD) is one of the effective methods to decompose high-dimensional fluid data into several significant modes of flowfields~\cite{taira2017modal}. Here, POD is a data-driven method which gives us the most significant and relevant structure in the data, and it exactly corresponds to principal component analysis and Karhunen-Lo\`{e}ve (KL) decomposition, where the decomposed modes are spatially orthogonal to each other. POD analysis for a discrete data matrix can be carried out by applying the singular value decomposition, as is often the case in the engineering fields.  A data matrix $\bm{X}\in \mathbb{R}^{n\times m}$ can be decomposed with POD in the following equation: $\bm{X}=\bm{U}\Sigma\bm{V}^\text{T}$. Here, columns of $\bm{U} \in \mathbb{R}^{n\times m}$ and $\bm{V} \in \mathbb{R}^{m\times m}$ are the spatial and temporal POD modes, respectively, and diagonal entries of $\Sigma \in \mathbb{R}^{m\times m}$ are the POD mode amplitudes.  Although there are several advanced data-driven methods, dynamic mode decomposition~\cite{schmid2010dynamic,Kutz2016}, empirical mode decomposition, and others which include efforts by the authors~\cite{nonomura2019extended}, this research is only based on POD, which is the most basic data-driven method for reduced-order modeling. 

If the data, such as flowfields, can be effectively expressed by a limited number of POD modes, limited sensors placed at appropriate positions give us the approximated full-state information. This effective observation might be one of the keys for flow control and flow prediction. This idea has been adopted by Manohar et al.~\cite{manohar2018data-driven}, and the sparse-sensor-placement algorithm has been developed and discussed. Previous studies show that a flowfield approximated by a small number of proper orthogonal decomposition modes can also be accurately reconstructed from a small number of sensors, by using a sparse-sensor-placement algorithm. The idea here is expressed by the following equation:
\begin{eqnarray}
 \bm{y}=\bm{Hx}\approx\bm{HU}_{r}\bm{z}=\bm{C}\bm{z}.
\label{eq:y_cx}
\end{eqnarray}
Here, $\bm{y}\in \mathbb{R}^p$, $\bm{H} \in \mathbb{R}^{p\times n}$, $\bm{x} \in \mathbb{R}^{n} $, $\bm{U}_r\in \mathbb{R}^{n\times r}$, $\bm{C}\in\mathbb{R}^{p\times r}$ and $\bm{z} \in \mathbb{R}^r$ are the observation vector, the sparse sensor location matrix, the full-state data vector, truncated $\bm{U}$, the measurement matrix ($\bm{C}=\bm{H}\bm{U}_r$) and the POD mode amplitude vector, respectively. In addition, $p$ and $r$ are the numbers of sensors and POD modes, respectively, and $n$ is the degree of freedom of the spatial POD modes. Equation (\ref{eq:y_cx}) can be solved as $\bm{z}=\bm{C}^{-1}\bm{y}$ in the case of $r=p$ and the optimization is equivalent to the maximization of the determinant of $\bm{C}$. The problem above is considered to be one of the sensor selection problems when $\bm{U}_r$ is assumed to be a sensor-candidate matrix. 
Thus far, this sensor selection problem has been solved by a convex approximation~\cite{joshi2009sensor} and a greedy algorithms~\cite{manohar2018data-driven}, where the greedy algorithm was shown to be much faster than the convex approximation algorithms. Table \ref{table:Computational_cost} summarizes the computational costs based on each calculation method: brute-force searching, convex approximation method, and greedy algorithm. The convex approximation method, which obtains suboptimal results~\cite{joshi2009sensor}, suffers from a high computational cost. A previous study~\cite{manohar2018data-driven} introduced a greedy algorithm based on the QR-discrete-empirical-interpolation method (QDEIM)~\cite{chaturantabut2010nonlinear,drmac2016new} when the number of sensors is the same as that of POD modes and its extension for the least squares problem when the number of sensors is greater than that of POD modes. Both convex approximation and greedy algorithms work pretty well for the sensor selection problems.

\begin{table}[ht]
\centering
\caption{Computational cost of sensor selection
 methods~\cite{manohar2018data-driven}.} 
\label{table:Computational_cost}
\begin{tabular}[t]{ll}
\hline
&Computational cost\\
\hline
Brute-force search& ${\frac{n!}{(n-p)!p!}} \sim {O(n^p)}$\\
Convex approximation method&${O(n^3)}$ per iteration.\\
Greedy method&${p = r}$ : ${O(nr^2)}$\\
\hline
\end{tabular}
\end{table}

There are several applications of a vector-measurement sensor, such as two components of velocity of particle image velocimetry, or simultaneous velocity, pressure, and temperature measurements used in weather forecasting, once the data are properly normalized. For instance, the authors have now been developing a sparse processing particle-image-velocimetry (PIV)-measurement system~\cite{kanda2019feasibility}. The real-time PIV measurement of the flowfield is required to perform active control of a high-speed flowfield in laboratory experiments. The velocity field is calculated from the cross-correlation coefficient for each interrogation window of the particle image in PIV measurement, but the number of windows that can be processed in a short time is limited. The amount of processing data is reduced and the flowfield is estimated by a limited number of selected windows located sparsely as in this study. 
The extension of the vector-measurement-sensor selection of the convex approximation has already been addressed in Sec. C, Chap. V of the original paper~\cite{joshi2009sensor}, while one of the greedy algorithms has not been proposed. The sensor selection of very high dimension with such a constraint should be resolved in reasonable time when the real-time measurement and flow control or flow prediction is to be conducted. Therefore, an extension of the greedy sparse sensor selection method to vector-measurement sensors is straightforwardly proposed and it is applied to PIV data to reconstruct the full state based on the information given by the sparse vector-measurement sensors.
In this study, we focus on proposing a greedy algorithm for the vector-measurement-sensor selection problem applied to the constant POD modes as the first step of a series of studies, whereas the assumption of constant POD modes is the same as that in the previous study~\cite{manohar2018data-driven}. The truncated POD provides the optimal rank-$r$ approximation to the data matrix, which can be exploited for reconstruction from limited in-situ measurements. Although the robustness against changes in POD modes should be considered because POD modes are not always constant in actual flowfields, this point is left for future study.

\section{Material and Methods}
\subsection{PIV Measurement}
PIV measurement for acquiring time-resolved data of flowfields around an airfoil was conducted previously~\cite{nankai2019linear}. Here, the experimental data are briefly explained. The wind tunnel testing was conducted in the Tohoku-university Basic Aerodynamic Research Wind Tunnel (T-BART) with a closed test section of 300 mm $\times$ 300 mm cross-section. The airfoil of the test model had an NACA0015 profile, the chord length and span width of which were 100 mm and 300 mm, respectively. The freestream velocity $U_\infty$ and attack angle of the airfoil $\alpha$ were set to be 10 m/s and 16 degrees, respectively. The chord Reynolds number was 6.4 $\times$ $10^4$. Time-resolved PIV measurement was conducted with a double-pulse laser. The time between pulses, the sampling rate, the particle image resolution, and the total number of image pairs were 100 $\mathrm{\mu}$s, 5000 Hz, 1024 $\times$ 1024 pixels, and $N=1000$, respectively. The tracer particles were 50\% aqueous solution of glycerin with estimated diameter of a few micrometers. The particle images were acquired by using the double pulse laser (LDY-300PIV, Litron) and a high-speed camera (SA-X2, Photron) which were synchronized to each other. 

\subsection{Previous Greedy Algorithm for Scalar-Measurement Sensors}
In the greedy algorithm based on QR decomposition for the scalar-measurement problem, the selection of the sensor position is based on maximizing the norm of the corresponding row vector of the sensor-candidate matrix. Let $\bm{W}$, $W_{ij}$, and $\bm{w}_i$ denote the sensor-candidate matrix, the $(i, j)$~th entry of $\bm{W}$, and the $i$~th row vector of $\bm{W}$, respectively. The $k$~th sensor is chosen from the $i$($=1,\dots,n$)~th sensor candidate where 
\begin{eqnarray}
i=\argmax_{i} \|\bm{w}_i\|^2_2.
\end{eqnarray}
Here, $\bm{w}_i=[W_{i,1} W_{i,2} \dots W_{i,r}]$, and $[W_{i,j}]=\bm{W}$ is initialized to be $\bm{U}_r$ for $p=r$ sensor conditions. Given the $i$ index, the $\bm{W}$ matrix is pivoted and QR decomposition is conducted, where this procedure resets the components of the selected sensor candidate to be the zero vector, and the already selected sensors are not selected again in the succeeding steps. After that, the next sensor is chosen for the remaining matrix. The algorithm for $p=r$ sensors is QDEIM~\cite{drmac2016new}.

The optimization is considered to be the maximization of the determinant of the $\bm{C}$ matrix to stably solve the $\bm{z}$ vector. The selection of the sensor position is based on maximizing the norm of the corresponding row vector of the sensor-candidate matrix. Although the round-off error increases, this procedure can be simply written as Gram-Schmidt orthogonalization with choosing the rows of the largest norm. Algorithm~\ref{alg:sch} shows the algorithm implemented in this study for the replacement of QR decomposition by the Gram-Schmidt procedure.
\begin{algorithm}
\caption{Greedy method for the scalar-measurement sensors }\label{alg:sch}
\begin{algorithmic}
\State Set sensor-candidate matrix $\bm{W}=\bm{U}_r$.
\State {$p=r$}
\For{ $k =1, \dots, p$ }
      \State $\bm{w}_{i}=[W_{i,1} W_{i,2} \dots W_{i,r}]$
      \State $i \leftarrow \argmax_i \|\bm{w}_i \|^2_2$
      \State $\bm{W} \leftarrow \bm{W}-\bm{W}\bm{w}_i^{\text{T}}\bm{w}_i/\|\bm{w}_i\|^2_2$
      \State $H_{k,i} = 1$
\EndFor
\State \textbf{return} $\bm{H}$
\end{algorithmic}
\end{algorithm}

\subsection{Proposed Greedy Method for Vector-Measurement Sensors Algorithm}
In the vector measurement, we consider the following equation:
\label{sec:propose}
\begin{eqnarray}
\bm{y}&=&\left[\begin{array}{cccc}\bm{H}&\bm{0}&\bm{0}&\bm{0}\nonumber\\
                                  \bm{0}&\bm{H}&\bm{0}&\bm{0}\nonumber\\
                                  \bm{0}&\bm{0}&\ddots&\bm{0}\\
                                 \bm{0}&\bm{0}&\bm{0}&\bm{H}\\\end{array}\right]
         \left[\begin{array}{c}\bm{U}_1 \\ \bm{U}_2 \\ \vdots \\ \bm{U}_s  \end{array}\right] \bm{z}\\
     &=& \qquad \qquad \bm{H}_s \qquad \qquad \bm{U}_r \quad \bm{z}\label{eq:vec}\\
     &=& \qquad \qquad \qquad  \bm{C} \qquad \qquad \bm{z}
\end{eqnarray}
Here, $\bm{U}_s \in \mathbb{R}^{\frac{n}{s} \times  r}$ is the $s$~th vector component of a sensor-candidate matrix, $\bm{H} \in \mathbb{R}^{p \times \frac{n}{s}}$ is the sensor location matrix for each vector component where, $s$ is the number of components of the measurement vector. Again, $p$ and $r$ are the numbers of sensors and POD modes, respectively, and  $n$ is the degree of freedom of the spatial POD modes including the different vector components. This arrangement of data is intentionally chosen with considering the situation when the data matrix of $\bm{X}=[\bm{X}_u^{\text{T}} \quad \bm{X}_v^{\text{T}}]^{\text{T}}$ is applied and the spatial POD modes of $\bm{X}$ are used as the sensor-candidate matrix, where $\bm{X}_u^{\text{T}}$ and $\bm{X}_v^{\text{T}}$ are data matrices of $x$ and $y$-velocity components in PIV data, respectively. This arrangement does not matter for the Gram-Schmidt procedure, but we recommend to reorder, the data as the data of the same vector-measurement sensor are gathered in successive rows when this algorithm is further straightforwardly extended by block-pivoting and the block-QR algorithm for eliminating round-off error. However, the latter extension is not addressed in this short note, for brevity.

Similar to the scalar-measurement, the next sensor can be chosen to maximize the determinant (submatrix volume in each step~\cite{manohar2018data-driven}) of the $\bm{C}$ matrix. Because the multiple ($s$) rows of the $\bm{U}_r$ matrix are chosen simultaneously by selecting one point in the vector version, the hypervolume of the selected row vectors is maximized, instead of the norm of the row vector in the scalar version.
Therefore, in the greedy algorithm for a vector measurement problem, the $k$~th sensor is chosen for the maximization of the hypervolume of the matrix from the $i$ ($=1,\dots,n/s$)~th sensor candidate as follows: 
\begin{eqnarray}
i=\argmax_{i} \|\bm{w}_{i} \wedge \bm{w}_{i+\frac{n}{s}} \wedge \dots \wedge \bm{w}_{i+\frac{n(s-1)}{s}} \|_2^2,
\end{eqnarray}
where $\wedge$ denotes the wedge product of the exterior algebra. The definition above recovers the original greedy method when $s=1$. This squared hypervolume ($J_i$ in the algorithm) can be simply computed by a Gram-Schmidt-like procedure by multiplying the norm of one row and removing its component from other selected rows in order, without considering the exterior algebra. After choosing the sensors, the components of selected row vectors are subtracted from the sensor-candidate matrix and then proceed to the next sensor placement selection. This procedure again avoids the selection of the same sensor location. 
The algorithm is summarized in Algorithm \ref{alg:vec}. The extension of the convex approximation method \cite{joshi2009sensor} for vector sensor placement is addressed in the original paper, and it is adopted in the present test case.

\begin{algorithm}
\caption{Greedy method for vector-measurement sensors}\label{alg:vec}
\begin{algorithmic}
   \State Set sensor-candidate matrix $\bm{U}_j$ for $j$th vector component measurement.
   \State $\bm{U}_r=\left[\begin{array}{cccc}\bm{U}_1^{\text{T}}&\bm{U}_2^{\text{T}}&\dots&\bm{U}_j^{\text{T}}\\\end{array}\right]^{\text{T}}$
   \State {$sp=r$}
   \State $\bm{W}=\bm{U}_r$
   \For{ $k =1, \dots, p$ }
      \For{ $i = 1, \dots, n$ }            
      \State $\bm{w}_{i} = \left[\begin{array}{cccc} W_{i,1}
      &W_{i,2} &\dots&W_{i,r} \end{array}\right]$
      \EndFor
      \For{ $i = 1, \dots, \frac{n}{s}$ }
          \State $J_i = 1$
          \State $\bm{\tilde{W}}= \bm{W}$
          \For{$j = 1, \dots, s$ }
            \State $J_i = J_i\|(\bm{\tilde{w}}_{i+\frac{n}{s}(j-1)}\|^2_2$
            \State $\bm{\tilde{W}} \leftarrow \bm{\tilde{W}}-\bm{\tilde{W}}(\bm{\tilde{w}}_{i+\frac{n}{s}(j-1)}^{\text{T}} \bm{\tilde{w}}_{i+\frac{n}{s}(j-1)}/\|\bm{\tilde{w}}_{i+\frac{n}{s}(j-1)}\|^2_2)$
          \EndFor
     \EndFor
      \State $i = \argmax_i J_i $
      \State $H_{k,i} = 1$      
      \For{ $j = 1, \dots, s $ }            
            \State $\bm{w}_{i+\frac{n}{s}(j-1)} = 
            \left[\begin{array}{cccc} W_{i+\frac{n}{s}(j-1),1} &V_{i+\frac{n}{s}(j-1),2} &\dots &W_{i+\frac{n}{s}(j-1),r}\end{array}\right]$      
            \State $\bm{W} \leftarrow \bm{W}-\bm{W} (\bm{w}_{i+\frac{n}{s}(j-1)}^{\text{T}}\bm{w}_{i+\frac{n}{s}(j-1)}/\|\bm{w}_{i+\frac{n}{s}(j-1)}\|^2_2)$
       \EndFor
   \EndFor\label{euclidendwhile}
   \State \textbf{return} $\bm{H}$
\end{algorithmic}
\end{algorithm}

\section{Results and Discussions}
\subsection{Random sensor problem}
\begin{figure}[htbp]
    \centering
    \includegraphics[scale=0.25]{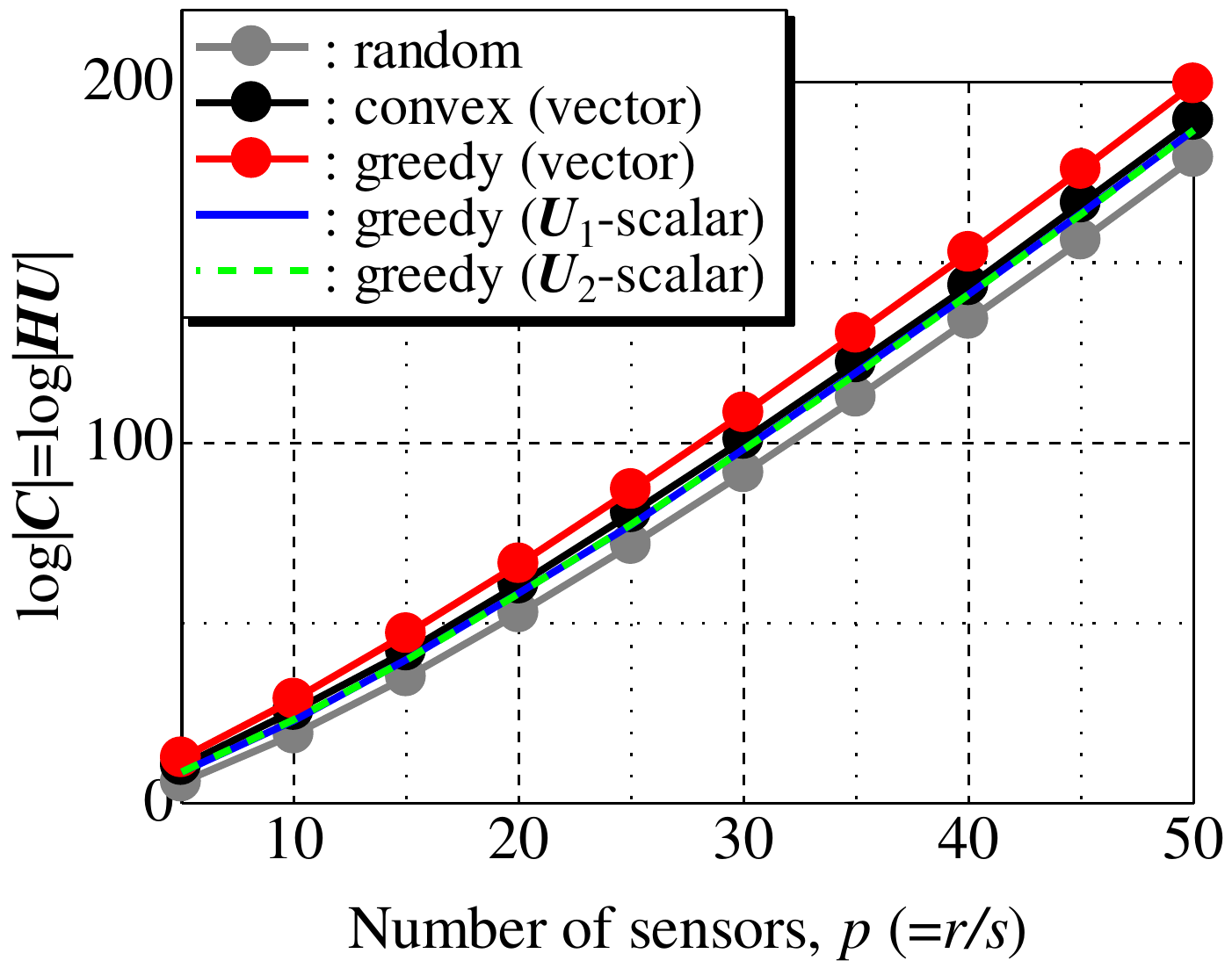}
    \caption{Relationship between the number of sensors and the logarithm of the determinant of $\bm{C}$ ($s=2$)}
    \label{fig:p.vs.log(det(C))}
\end{figure}

Numerical experiments are conducted and the proposed algorithm is validated for the vector-measurement sensor placement problem in the multiple components of the measurement vector ($s=2$). The random sensor-candidate matrices, $\bm{U}_1\in \mathbb{R}^{1000{\times}r}$  and $\bm{U}_2 \in \mathbb{R}^{1000{\times}r}$, were set, where each component of the matrices is given by the Gaussian distribution of $\mathcal{N}(0,1)$. Therefore, the sensor-candidate matrix $\bm{U}_r$ is expressed as $\bm{U}_r$=[$\bm{U}_1^\text{T}\quad \bm{U}_2^\text{T}$]$^\text{T}$ in this validation. The sparse sensor location matrix is calculated based on each method: random selection, convex approximation method, and greedy algorithms of vector-measurement and scalar-measurement sensors, respectively. The greedy algorithm of scalar-measurement sensors selects $r/2$ sensors based on one sensor candidate matrix as opposed to the greedy algorithm of vector-measurement sensors which selects $r/2$ sensors based on both sensor-candidate matrices. After selecting sensors, the logarithm of the determinant of $\bm{C}=\bm{H}$$\bm{U}_r$ is calculated using both sensor-candidate matrices. Therefore, the other components of the vector-measurement sensors are randomly selected because they are not considered in the process of sensor selection. Of course, the greedy algorithm of scalar measurement is not considered to succeed to select sensors due to a lack of proper treatment in the other components of a vector-measurement sensor-candidate matrix. Fig.~\ref{fig:p.vs.log(det(C))} shows the relationship between the number of sensors and the logarithm of the determinant of $\bm{C}=\bm{H}$$\bm{U}_r$. Here, greedy (vector), greedy ($\bm{U}_1$-scalar), and greedy ($\bm{U}_2$-scalar) in Fig.~\ref{fig:p.vs.log(det(C))} are calculation results based on the greedy algorithm for vector-measurement sensors using $\bm{U}_1$ and $\bm{U}_2$, that for scalar-measurement sensors using $\bm{U}_1$, and that for scalar-measurement sensors using $\bm{U}_2$, respectively. All plots are average values of 100 calculations changing $\bm{U}_1$ and $\bm{U}_2$ as a normal random number every calculation. Fig.~\ref{fig:p.vs.log(det(C))} shows that the values of $\bm{C}$ obtained by the vector greedy algorithm are the highest value in all calculation results in all conditions. Because both greedy ($\bm{U}_1$-scalar) and greedy ($\bm{U}_2$-scalar) calculate sensor placements using only $\bm{U}_1$ or $\bm{U}_2$ although the sensor-candidate matrix $\bm{U}_r$ consists of $\bm{U}_1$ and $\bm{U}_2$, half of the sensor components of the greedy method for the scalar-measurement sensors has almost the same quality as the sensor placement obtained by the random selection. Therefore, all the results of the logarithm of the determinant of the greedy method for scalar-measurement sensors are in between those of the greedy method for vector-measurement sensors and of the random selection. The optimization is considered to conduct the maximization of the determinant of the $\bm{C}$ matrix to stably solve $\bm{z}$ vector as explained in Secs. \ref{sec:intro} and \ref{sec:propose}. Therefore, the greedy (vector) algorithm is more effective for sparse sensor placement than the greedy (scalar) and convex approximation in multiple components of the measurement ($s=2$).

\subsection{PIV measurement}
The reduced-order PIV data are reconstructed by sparse sensors that are chosen by several methods. Further, the PIV data for flows around airfoils are adopted, and the number of POD modes is predetermined. Here, the number of POD modes $r$ increase as that of sensors $p$ increases ($r=ps$). The extended convex approximation method is called the convex (vector) method, for brevity. The greedy algorithm for vector-measurement and scalar-measurement sensors of $u$ and $v$ fields is applied. These methods are called the greedy (vector), greedy ($u$-scalar), and greedy ($v$-scalar) methods for brevity.

\begin{figure}[htbp]
    \centering
    \includegraphics[scale=0.25]{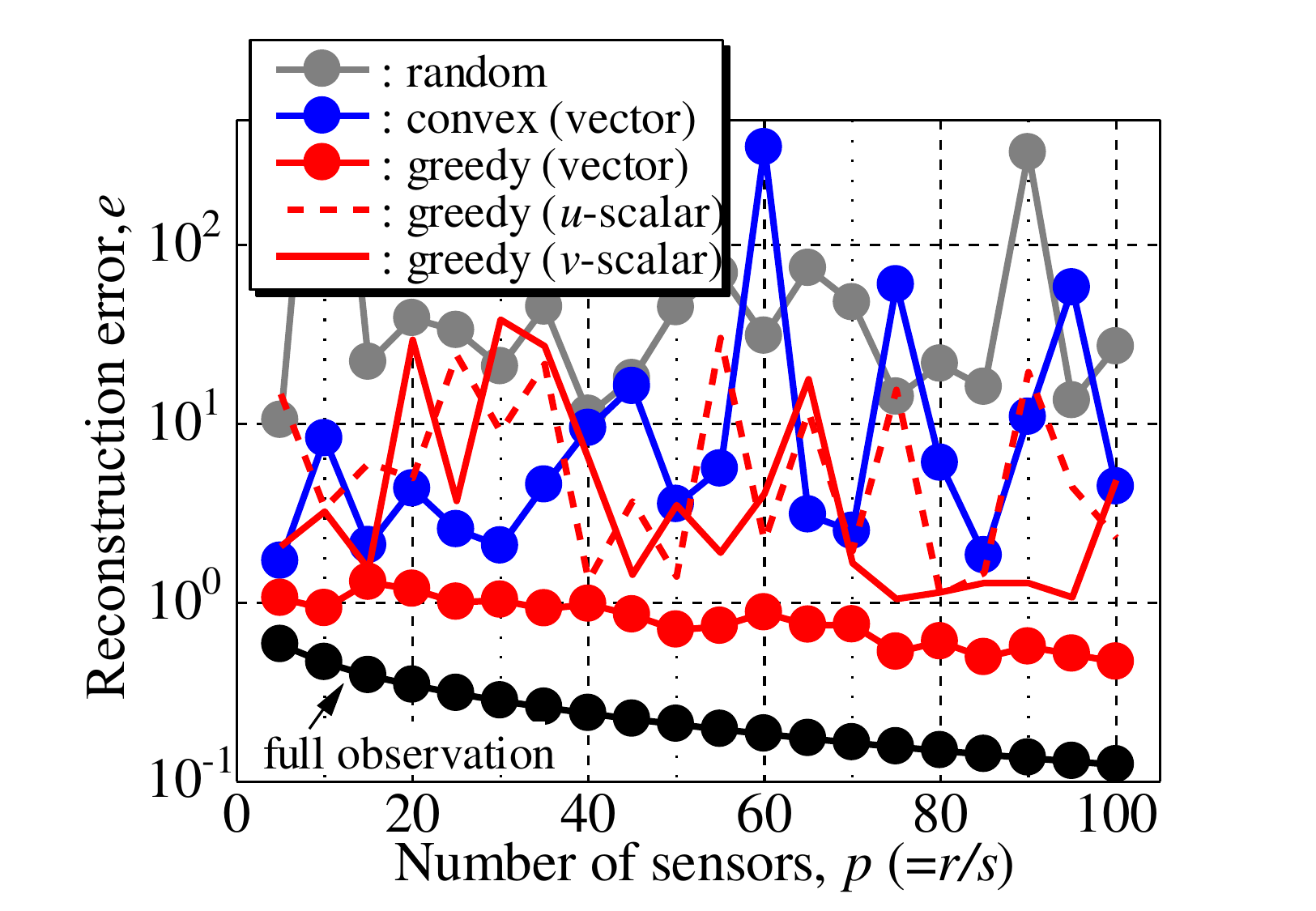}
    \caption{The relationship between the number of sensors and the reconstruction error}
    \label{fig:error}
\end{figure}

Fig.~\ref{fig:error} shows the relationship between the reconstruction error $e$ and the number of sensors $p$, obtained by full observation (a black solid line with closed circles), random selection (a gray solid line with closed circles), convex (vector) (a blue solid line with closed circles), greedy (vector) (a red solid line with closed circles), greedy ($u$-scalar) (a red dotted line), and greedy ($v$-scalar) methods (a red solid line). The reconstruction error $e$ is introduced for quantitative evaluation of flowfield reconstruction:
\begin{eqnarray}
  e = \frac{\sum_{j=1}^N\|\bm{x}_{j}-\hat{\bm{x}}_{j}\|^2_2}{\sum_{j=1}^N\|\bm{x}_{j}\|^2_2}
    \label{eq:error}
\end{eqnarray}
where, $\hat{\bm{x}}$ is the estimated data vector by using a sparse sensor measurement and competed as follows: $\hat{\bm{x}}=\bm{U}_{r}\bm{z}=\bm{U}_{r}\bm{C}^{-1}\bm{y}$. Here, the subscript $j$~denotes the quantity of $j$~th time step and $N$ represents the number of tested data. In the full observation, the estimated data vector is computed as follows: $\hat{\bm{x}}=\bm{U}_{r}\bm{U}_{r}^{\rm{T}}\bm{x}$.
The reconstruction error 
in the numerator of (\ref{eq:error}) is equivalent to the difference between the full-state observation and that reconstructed using sparse sensors, and the minimization of the reconstruction error relates to the experiment design minimizing the volume of the resulting confidence ellipsoid~\cite{joshi2009sensor}. It is difficult to estimate the accurate flowfield from the noisy flowfield by using the limited number of sensors. Therefore, the reconstructed errors of the methods excluding the full observation are sometimes over unity as shown in Fig.~\ref{fig:error}. The reconstruction error of full observation and greedy (vector) decreases as the numbers of POD modes and sensors increase. On the other hand, the reconstruction errors of the random selection, the convex (vector) approximation method, and both greedy ($u$-scalar and $v$-scalar) algorithms do not decrease as the number of sensors increases. The greedy ($u$-scalar and $v$-scalar) algorithms determine the sensors based on either the $u$ or $v$ field, but the counter component of the sensors of the $v$ or $u$ field is also used for the reconstruction. This counter component of the sensors is similar to those selected randomly because this counter component is not considered in the process of the sensor selection. Therefore, the proposed method works better than choosing half of the sensors using greedy ($u$-scalar and $v$-scalar) algorithms.
Because the convex approximation method does not always select optimal sensors, the reconstruction error of the convex (vector) approximation does not decrease as the numbers of sensors and POD modes increase, as shown in Fig.~\ref{fig:error}. 
Therefore, the proposed method is more efficient in terms of the reconstruction error and the computational cost as presented in Table \ref{table:Computational_cost}.

\section{Conclusions}
The greedy method extended to vector-measurement-sensor problems, such as the sensor placement problem in PIV data of fluid dynamic fields, is introduced and investigated in this paper. The sensor selection problem is solved by the convex approximation method for vector sensors and greedy methods for both scalar-measurement and vector-measurement sensors, where greedy algorithms are shown to be much faster than the convex approximation algorithms. The proposed method was validated using the random sensor problem before being adopted for PIV data, and the calculation result shows that the proposed method is more effective for the sparse vector-measurement sensor placement than the greedy method for scalar-measurement sensors and the convex approximation method for vector-measurement sensors. The calculation results of PIV data show that the reconstruction error of the greedy method for the vector-measurement sensors is smaller than other methods in the $r=ps$ condition. Therefore, the greedy method extended to the vector-measurement-sensor problem is illustrated to be more effective for sparse sensing than other methods.

\section*{Acknowledgements}
This work was supported by JST CREST, JPMJCR1763, Japan.
\normalsize
\bibliographystyle{IEEEtran}
\bibliography{xaerolab}

\begin{thebibliography}{10}
\providecommand{\url}[1]{#1}
\csname url@samestyle\endcsname
\providecommand{\newblock}{\relax}
\providecommand{\bibinfo}[2]{#2}
\providecommand{\BIBentrySTDinterwordspacing}{\spaceskip=0pt\relax}
\providecommand{\BIBentryALTinterwordstretchfactor}{4}
\providecommand{\BIBentryALTinterwordspacing}{\spaceskip=\fontdimen2\font plus
\BIBentryALTinterwordstretchfactor\fontdimen3\font minus
  \fontdimen4\font\relax}
\providecommand{\BIBforeignlanguage}[2]{{%
\expandafter\ifx\csname l@#1\endcsname\relax
\typeout{** WARNING: IEEEtran.bst: No hyphenation pattern has been}%
\typeout{** loaded for the language `#1'. Using the pattern for}%
\typeout{** the default language instead.}%
\else
\language=\csname l@#1\endcsname
\fi
#2}}
\providecommand{\BIBdecl}{\relax}
\BIBdecl

\bibitem{taira2017modal}
K.~Taira, S.~L. Brunton, S.~T. Dawson, C.~W. Rowley, T.~Colonius, B.~J. McKeon,
  O.~T. Schmidt, S.~Gordeyev, V.~Theofilis, and L.~S. Ukeiley, ``Modal analysis
  of fluid flows: An overview,'' \emph{AIAA Journal}, pp. 4013--4041, 2017.

\bibitem{schmid2010dynamic}
P.~J. Schmid, ``Dynamic mode decomposition of numerical and experimental
  data,'' \emph{Journal of Fluid Mechanics}, vol. 656, no. July 2010, pp.
  5--28, 2010.

\bibitem{Kutz2016}
J.~N. Kutz, S.~L. Brunton, B.~W. Brunton, and J.~L. Proctor, \emph{Dynamic mode
  decomposition: data-driven modeling of complex systems}.\hskip 1em plus 0.5em
  minus 0.4em\relax SIAM, 2016, vol. 149.

\bibitem{nonomura2019extended}
T.~Nonomura, H.~Shibata, and R.~Takaki, ``Extended-kalman-filter-based dynamic
  mode decomposition for simultaneous system identification and denoising,''
  \emph{PloS one}, vol.~14, p. e0209836, 2019.

\bibitem{manohar2018data-driven}
K.~{Manohar}, B.~W. {Brunton}, J.~N. {Kutz}, and S.~L. {Brunton}, ``Data-driven
  sparse sensor placement for reconstruction: Demonstrating the benefits of
  exploiting known patterns,'' \emph{IEEE Control Systems Magazine}, vol.~38,
  no.~3, pp. 63--86, June 2018.

\bibitem{joshi2009sensor}
S.~Joshi and S.~Boyd, ``Sensor selection via convex optimization,'' \emph{IEEE
  Transactions on Signal Processing}, vol.~57, no.~2, pp. 451--462, 2009.

\bibitem{chaturantabut2010nonlinear}
S.~Chaturantabut and D.~C. Sorensen, ``Nonlinear model reduction via discrete
  empirical interpolation,'' \emph{SIAM Journal on Scientific Computing},
  vol.~32, no.~5, pp. 2737--2764, 2010.

\bibitem{drmac2016new}
Z.~Drmac and S.~Gugercin, ``A new selection operator for the discrete empirical
  interpolation method---improved a priori error bound and extensions,''
  \emph{SIAM Journal on Scientific Computing}, vol.~38, no.~2, pp. A631--A648,
  2016.

\bibitem{kanda2019feasibility}
N.~Kanda, K.~Nankai, Y.~Saito, T.~Nonomura, and K.~Asai, ``Feasibility study on
  sparse processing particle image velocimetry,'' \emph{Bulletin of the
  American Physical Society}, 2019.

\bibitem{nankai2019linear}
K.~Nankai, Y.~Ozawa, T.~Nonomura, and K.~Asai, ``Linear reduced-order model
  based on piv data of flow field around airfoil,'' \emph{TRANSACTIONS OF THE
  JAPAN SOCIETY FOR AERONAUTICAL AND SPACE SCIENCES}, vol.~62, no.~4, pp.
  227--235, 2019.

\end{thebibliography}

\end{document}